\documentclass[twocolumn]{jpsj2} 

\title{Infrared study of spin crossover Fe-picolylamine complex}

\author{%
Hidekazu \textsc{Okamura}\thanks{E-mail: okamura@phys.sci.kobe-u.ac.jp}, 
Masato \textsc{Matsubara}, 
Takeshi \textsc{Tayagaki}$^1$, 
Koichiro \textsc{Tanaka}$^{1}$, \\
Yuka \textsc{Ikemoto}$^2$, 
Hiroaki \textsc{Kimura}$^2$ 
Taro \textsc{Moriwaki}$^2$ and 
Takao \textsc{Nanba}, 
}

\inst{%
Graduate School of Science and Technology, Kobe University, 
Kobe 657-8501 
\\
$^1$Department of Physics, Graduate School of Science, 
Kyoto University, Kyoto 606-8502 
\\
$^2$Japan Synchrotron Radiation Research Institute, 
Sayo 679-5198.
}

\recdate{\today}

\abst{%
Infrared (IR) absorption spectroscopy has been used to 
probe the evolution of microscopic vibrational states upon 
the temperature- and photo-induced spin crossovers in 
[Fe(2-picolylamine)$_3$]Cl$_2$EtOH (Fe-pic).     
To overcome the small sizes and the strong IR absorption of 
the crystal samples used, an IR synchrotron radiation source and 
an IR microscope have been used.   
The obtained IR spectra of Fe-pic show large changes between 
high-spin and low-spin states for both the temperature- 
and the photo-induced spin crossovers.      
Although the spectra in the temperature- and photo-induced 
high-spin states are relatively similar to each other, 
they show distinct differences below 750 cm$^{-1}$.    
This demonstrates that the photo-induced high-spin state 
involves microscopically different characters from 
those of the temperature-induced high-spin state.   
The results are discussed in terms of local pressure and 
structural deformations within the picolylamine ligands, 
and in terms of their possible relevance to the development of 
macroscopic photo-induced phase in Fe-pic.  
}

\kword{
spin crossover complex, photo-induced phase transition, 
infrared spectroscopy, infrared synchrotron radiation
}

\begin{document}
\maketitle

\section{Introduction}
Recently, photo-induced spin crossover (SC) phenomena have attracted 
a lot of interest in the condensed matter physics community.    
Various compounds have been shown to exhibit photo-induced 
SC, many of which are organic complexes containing 
transition metal elements.      
[Fe(2-pic)$_3$]Cl$_2$EtOH (2-pic = 2-picolylamine = 
2-aminomethyl pyridine, EtOH = ethanol), 
hereafter referred to as Fe-pic, is one of such spin-crossover 
compounds.\cite{Fe-pic1,Fe-pic2,time,koshihara,taya-PRL}    
As illustrated in Fig.~1(a), in Fe-pic an Fe$^{2+}$ ion is coordinated 
by three 2-picolylamine ligands, where Fe$^{2+}$ is octahedrally 
surrounded by nearest neighbor N atoms.   A picolylamine 
molecule is sketched in Fig.~1(b).   Four formula units of 
[Fe(2-pic)$_3$]Cl$_2$EtOH are contained in a unit cell.     
Two different electron configurations are possible for Fe$^{2+}$, 
depending on the Hund coupling among the 3$d$ electrons 
and the crystal field splitting, as sketched in Fig.1~(c).   
If the former is greater, 
Fe$^{2+}$ takes the total spin $S$=2 (``high-spin'' state), 
and otherwise it takes $S$=0 (``low-spin'' state), where the 
Fe-N bond length is larger in $S$=2 state.       
Fe-pic undergoes a SC from $S$=2 to 0 upon cooling 
through T$_c$ $\sim$~115~K as shown in Fig.~1(d), where 
$\gamma_{HS}$, the fraction of high-spin Fe$^{2+}$ ions 
deduced from the measured magnetic susceptibility, is plotted 
as a function of temperature.\cite{taya-thesis}      
Far below $T_c$, the conversion to $S$=0 
is almost complete.    The SC occurs in a 
temperature width of about 30~K.   (The plateau-like 
structure seen in the middle of crossover in $\gamma_{HS}$ 
is of great interest,\cite{2step} but not discussed in this work.)      
Below $T_c$, a photoexcitation can cause 
a crossover back to $S$=2.     At low temperatures, this 
photo-induced $S$=2 state has a long lifetime even after 
the photoexcitation is turned off, reaching a few hours 
at 10~K, .     
Hereafter, we refer to these three characteristic states 
of Fe-pic as follows: the high-temperature, high-spin ($S$=2) 
state above $T_c$ as ``HTHS'', the low-temperature, low-spin 
($S$=0) state below $T_c$ as ``LTLS'', and the photo-induced, 
high-spin ($S$=2) state below $T_c$ as ``PIHS''.

A stabilization of high-spin state under photoexcitation 
at low temperatures has been found for many different 
SC compounds, and referred to as the ``light-induced excited 
spin-state trapping''.    In the case of Fe-pic, the 
development of PIHS has a 
characteristic threshold for the photoexcitation intensity, 
and an incubation time exists between the start of 
photoexcitation and the appearance of PIHS.\cite{koshihara}   
In addition, the relaxation of PIHS to LTLS is strongly 
non-exponential.\cite{time}     
These characteristic features strongly suggest that 
the appearance of PIHS in Fe-pic 
involves a collective interaction among the 
Fe$^{2+}$ ions.    Because of this, it has been 
suggested that the appearance of PIHS in Fe-pic should be a 
transition to a new macroscopic phase under photoexcitation, 
referred to as the 
{\it ``photo-induced phase transition''.}\cite{taya-PRL}     
Its general understanding is the following: After the 
photoexcitation of low-spin Fe$^{2+}$ ions to high-energy 
states, some of them relax to the high-spin state, instead of 
relaxing directly to the initial low-spin state.     
Since the relaxation to the high-spin state involves a large 
lattice deformation (expansion), it may have a long lifetime 
when the thermal fluctuation is quenched at low temperatures.     
Because of this, the density of high-spin Fe$^{2+}$ ions 
increases with time.     When it reaches a critical value, 
the local deformations at different Fe$^{2+}$ sites act 
collectively, leading to the development of PIHS over a 
macroscopic volume.     However, the microscopic nature of 
the PIHS has not been understood well.    For example, 
there has been a lot of discussion whether or not the PIHS 
and the HTHS are the identical $S$=2 states.    
In addition, it is not well known how the local deformations 
at the FeN$_6$ clusters lead to a long-range, collective 
interaction among them required for the development of PIHS.

To answer the above questions, it is important to obtain 
microscopic information about the electronic and vibrational 
states of Fe-pic in the three states.    For this purpose, 
both Raman and infrared (IR) spectroscopies have been 
performed.\cite{taya-PRL,taya-JL,taya-PT}    
Raman spectroscopy has the advantage of being able to use 
resonance effects by tuning the incident photon energy, 
which is particularly useful in studying transition-metal 
complexes such as Fe-pic.     
Indeed, the Raman studies of Fe-pic have provided 
much insight into the microscopic properties of 
Fe-pic.     
With the IR spectroscopy, one can use very low intensity 
and small photon-energy for the incident beam.  
This is also useful for the study of photo-induced SC, 
since one can separate the photoexcitation 
source and the probe IR source.     Previous IR works of 
Fe-pic\cite{taya-PRL,taya-JL,taya-PT} had been performed 
on powder samples pressed into KBr pellets.  
This was due to the small sizes (typically less than 1 mm) of the 
crystalline Fe-pic, and due to its very strong absorption 
in the mid-IR range.    Fe-pic is a soft material, and 
is very easily oxidized.    Hence with powder samples, 
there is a risk of obtaining 
extrinsic effects arising from surface oxidization, stress 
and strain caused by pressing, and also from size effects 
when using long-wavelength IR radiation.  (Wave numbers below 
2000~cm$^{-1}$ correspond to wavelengths longer than 5~$\mu$m, 
which may be well comparable to the powder sizes.)    
Hence it is strongly desired to use crystalline samples to 
obtain the intrinsic bulk information.  

In this work, we have studied in detail the IR absorption 
of Fe-pic using bulk crystal samples at various temperatures 
and under photoexcitation, to obtain intrinsic information 
about the microscopic vibrational states of Fe-pic.     
To overcome the above-mentioned difficulties with using 
crystal samples, we have used a synchrotron radiation 
IR source and an IR microscope.     
Many absorption lines have shown marked changes in the 
intensity and/or the peak position upon the temperature- 
and photo-induced SC's.     
In addition to the strong absorption lines previously observed 
with powder samples, we have newly found many weak lines that 
also show marked variations upon SC's.     
Although the absorption spectra are relatively similar between 
HTHS and PIHS, there are clearly different 
spectral features below 720~cm$^{-1}$.   Namely, HTHS and PIHS 
involve microscopically different nature.    We will discuss the 
implications of this result.

\section{Experimental}
The crystal samples of Fe-pic used in this work 
were grown by an evaporation method.      
Plate-shaped samples with a typical size of 
0.5$\times$0.5$\times$0.1~mm$^3$ were used.    
The IR spectroscopy was 
performed using a synchrotron radiation source and an 
IR microscope at the beam line BL43IR of 
SPring-8.\cite{kimura1,kimura2}    
Since the synchrotron radiation is emitted from an electron 
beam, the effective source size is much smaller than the 
conventional source.      
When focused onto the sample through the microscope, 
a spot diameter as small as 15~$\mu$m was obtained without 
using any aperture or pin-hole.   To achieve the same spot 
size with a conventional IR source, it would be necessary to 
place a small pin-hole within the optical path, which should 
result in much smaller intensity.   With such low intensity 
of incident IR beam, it would be impossible to study the 
transmission spectra of Fe-pic, which is strongly absorbing 
in the mid-IR range.   To cool the sample a liquid He 
continuous flow cryostat was used, and to record the IR 
spectra a Fourier transform IR interferometer and a HgCdTe 
detector were also used.   

\section{Results and Discussion}
In Section 3.1 the measured IR spectra of Fe-pic are presented as 
a function of temperature on going from HTHS to LTLS.    
In Section 3.2, the IR spectra after photoexcitation are 
presented, and the spectra in the three states are compared to 
one another.     
In Section 3.3, first the observed IR lines are assigned to 
specific vibration modes of Fe-pic, and then the discussion 
will be given.

\subsection{IR spectra of Fe-pic as a function of 
temperature}
%
%
Figure 2 shows the optical density (OD) spectra of Fe-pic at 
temperatures above and below $T_c$ recorded without photoexcitation.   
OD is a measure of absorption, given as -log[$I(\nu)/I_0(\nu)$], 
where $\nu$ is the wave number, $I(\nu)$ is the transmission 
spectrum of the sample and $I_0(\nu)$ is the reference spectrum 
recorded without sample.    
The detection limit for a weak transmission was about OD=2.7 when 
the accumulation time was 2~min, and Fig.~2 shows the spectra below 
OD=2.5 only.      A large number of absorption lines are 
observed in Fig.~2, which are basically due to intramolecular 
vibration modes of picolylamine, ethanol, and FeN$_6$ clusters.    
Intermolecular or lattice vibrations should have much lower 
frequencies, since they involve vibrations of the entire 
Fe-picolylamine complex.     With decreasing 
temperature from above to below $T_c$, hence going from HTHS 
($S$=2) to LTLS ($S$=0), some of the lines show 
large variations in their intensities and/or frequencies.   
In Fig.~2, some of such absorption lines are indicated by 
the shaded areas.   Note that those at 600-750~cm$^{-1}$ 
show particularly large temperature variations.    
In addition, weak but characteristic lines have been 
observed at 2150-2400~cm$^{-1}$, as shown in Fig.~3.    
These lines also show very large intensity changes 
with decreasing temperature.     
The strong absorption lines seen in Fig.~2 had 
been already observed in the previous work based on powder 
samples,\cite{taya-JL,taya-PT} but many of the weak lines 
in Figs.~2 and 3 had been overlooked.

Figure 4 shows the temperature dependence of the intensity 
for several absorption lines seen in Figs.~2 and 3.   
For each curve, the intensity integrated over the wave number range 
indicated in the figure has been normalized by the difference 
between 40 and 200~K.      It is very clear that the 
intensity variations during the SC closely follow that of 
$\gamma_{HS}$, shown in Fig.~1(d).    
In particular, the temperature width over which the 
intensity changes is as narrow as that in $\gamma_{HS}$.   
Apparently, the vibration modes giving rise to these 
absorption lines are strongly coupled with the spin state 
of Fe$^{2+}$.   In the previous work with powder 
samples,\cite{taya-PT} the intensity change of an IR 
absorption line during the SC was observed much broader 
than that in $\gamma_{HS}$.       
Hence the present data based on the crystal samples 
appear to provide more intrinsic information regarding the 
microscopic vibrational states in Fe-pic.

\subsection{IR spectra in the PIHS, and comparison with 
those in the other two states}  
%
%
Figure 5 shows the IR absorption spectra of Fe-pic in LTLS 
and PIHS at 10~K, which were recorded before and after a 
photoexcitation for 5 min, respectively.      Some of 
the observed IR lines have been labeled by the letters 
$a$ to $i$ in Fig.~5.     
The photoexcitation was made by a white light from a 
tungsten lamp with a power density of $\sim$ 1~mW/mm$^2$, 
and it was turned off before the data accumulation.       
The white light had a broad spectral distribution from visible 
to near-IR, which covered a wide photon energy range below the 
absorption edge of Fe-pic.     
Hence the excitation light had a large penetration 
depth and the photoexcitation was made uniformly within 
the sample volume.      
The PIHS of Fe-pic has a long lifetime at 10~K, and indeed 
the IR absorption spectra measured immediately after and 
5 min after the turn-off of the excitation source were 
nearly identical.     
In Fig.~5, many of the lines exhibit marked variations upon 
photoexcitation.   Particularly strong variations are evident 
for the lines below 750~cm$^{-1}$, those around 1200~cm$^{-1}$ 
and 1600 cm$^{-1}$, and those between 2200 and 2400~cm$^{-1}$.   
Since the sample temperature is the same for the two spectra, 
these spectral variations are purely due to changes in the 
microscopic vibrational states caused 
by the photo-induced SC.      

Figure 6 compares the IR spectra in HTHS (140~K) and 
PIHS (10~K).  The spectrum in PIHS is the same as that in Fig.~5, 
and the line labels are also the same as those in Fig.~5.    
Above 750~cm$^{-1}$, the two spectra in Fig.~6 seem to have 
no fundamental differences, taking into account the large 
temperature difference (10 and 140~K).      
Below 750 cm$^{-1}$, however, the two spectra exhibit 
remarkable differences:    
First, the line $b$ is present in PIHS, but not in HTHS.    
Second, the lines $b'$, $c$ and $b''$ are apparently 
shifted toward higher frequency in PIHS.     
Since no such shifts are observed above 750~cm$^{-1}$, 
the shifts of the $b'$, $c$ and $b''$ lines should not be due 
to thermal expansion of the crystal as a whole, but 
due to a local deformation of specific fragments within 
the crystal structure of Fe-pic.

\subsection{Line assignments and discussion}   
To assign the observed absorption lines to specific vibration 
modes of Fe-pic, we first compared the observed spectra to 
the published IR spectra of 2-picolylamine and 
ethanol.\cite{aldrich}     
  To distinguish the absorption lines related with the 
aminomethyl base from those related with the pyridine ring 
[see Fig.~1(a) and (b)], we compared the published spectra 
of 2-picolylamine with those of 2-amino-pyridine, 
2-methyl-pyridine, and pyridine.\cite{aldrich}      
To assign the IR lines to specific vibration modes, we 
followed the standard guidelines for the IR line assignment 
of organic compounds.\cite{IR-1,IR-2}     
Characteristic vibration modes of a specific fragment, for 
example the bending vibration modes of the pyridine ring when 
it is 2-substituted as in the case of 2-picolylamine, have 
been known to appear in characteristic frequency 
bands.\cite{IR-1,IR-2}     It was possible to assign most of 
the observed IR lines to specific vibration modes by these 
procedures.    
Table~I summarizes the results of the assignment.   
The lines above 750~cm$^{-1}$ are due to the C-H deformation 
modes and the skeletal stretching modes of the picolylamine.   
For the bands $i$ and $j$, the C-H deformation modes within 
the picolylamine and the ethanol, and the C-N stretching modes 
of the aminomethyl base overlap.\cite{IR-1,IR-2}     
   In contrast, the lines below 750~cm$^{-1}$ 
result from the skeletal deformation (bending) of picolylamine, 
in addition to the low-frequency C-H deformation modes.   
The lines $b$, $b'$ and 
$b''$ in Fig.~5, which show particularly large variations with 
the SC, could not be assigned with the above procedures only.   
They were assigned to the skeletal bending modes of 
$-$C$-$C$-$N$-$ in the aminomethyl base as explained below,

Detailed structure studies of Fe-pic using X-ray diffraction 
(XRD) have been reported in HTHS and LTLS\cite{mikami}, 
and also in PIHS.\cite{huby}     They have shown the following 
results:   The lattice constants decrease by $\sim$~1\% 
from HTHS (150~K) to LTLS (80~K), and increase by $\sim$~1\% from 
LTLS to PIHS at 10~K.     
On the other hand, the average Fe-N distance decreases by 
$\sim$~8~\% from HTHS (150~K) to LTLS (90~K), 
and increases by $\sim$~9~\% from LTLS to PIHS at 
10~K.\cite{huby} 
Namely, the contraction and the expansion of FeN$_6$ 
upon the temperature- and photo-induced SC's 
are much greater than those of the unit cell.  
This strong contraction/expansion of FeN$_6$ 
is expected to produce a physical pressure 
on the picolylamine molecules, which bridge two of the 
six N atoms in the FeN$_6$ cluster.    This will certainly 
result in a deformation of the picolylamine molecules.  
Hence the vibration modes of picolylamine actually 
involving such deformation, {\it i.e.}, the skeletal 
vibration modes, are expected to be significantly 
affected upon the SC; the lines $b$, $b'$ and $b''$ 
are likely to result from such vibration modes.  
The XRD results\cite{huby} have also shown that HTHS and 
PIHS are quite similar in terms of the {\it average} 
crystal structure, with the same space group and very 
close values of lattice constants and average Fe-N distance.    
Regarding the {\it local} structure, however, 
an important difference has been found between 
PIHS and HTHS by the X-ray absorption near-edge 
structure (XANES) and extended X-ray absorption fine 
structure (EXAFS) experiments.\cite{XAS}     
Although XANES and EXAFS cannot directly determine the 
overall crystal structure, they can sensitively probe the 
local environment of Fe$^{2+}$ ions.  
Their main results are the following:\cite{XAS} 
\\
\noindent (i) The local symmetry at Fe$^{2+}$ is $O_h$ in 
all the three states, {i.e.}, it is kept unchanged in 
both the temperature- and the photo-induced SC's.  
\\
\noindent (ii) The distribution in the distance between 
Fe$^{2+}$ and the next-nearest-neighbor (nnn) C atoms is 
slightly different between HTHS and PIHS.  
\\
\noindent   The result (i) shows the absence of 
Jahn-Teller effect in Fe-pic.        
Since all the nnn C stoms are contained in the ligand 
picolylamine, the result (ii) strongly suggests 
that the picolylamine molecules in PIHS are slightly deformed 
compared with those in HTHS.   Such deformation is probably a 
consequence of the physical pressure created by the expansion 
of FeN$_6$ clusters, 
and should also be related with the appearance of the $b$ line 
in PIHS and the shifts of the $b'$ and $b''$ lines between 
PIHS and HTHS.   
Based on these results from XRD, XANES, and EXAFS, 
we have assigned the peaks $b$, $b'$ and $b''$ 
to the skeletal bending of the $-$C$-$C$-$N part in the 
aminomethyl base.     It is known\cite{IR-1,IR-2} that 
the C-N stretching modes appear in the range 1050-1300~cm$^{-1}$.   
Hence the range 600-700 cm$^{-1}$ is reasonable for 
the corresponding bending vibrations.

Finally, the $l$ lines in Fig.~5, observed at 
2300-2400~cm$^{-1}$ and showing very large variations upon 
the SC, probably result from overtones and/or coupled 
modes of lower-frequency lines because no characteristic 
vibrations are expected in this range for picolylamine and ethanol 
molecules.\cite{aldrich,IR-1,IR-2}    However, it is unclear 
which of the lower-frequency lines have actually resulted in 
the $l$ lines.        In spite of the small intensities, 
their spectral changes due to temperature- and photo-induced 
SC's are remarkably clear as shown in Figs. 3 and 5.     
Hence, the $l$ lines may be due to coupled modes including 
FeN$_6$ cluster vibrations.    
Note that the line energies (approximately 0.3~eV) are 
too small to be crystal field levels of Fe$^{2+}$.

The present results based on crystal samples of Fe-pic have 
shown that the microscopic vibration states undergo very 
large changes upon the temperature- and photo-induced SC's.     
These results are reasonable since a change of approximately 
8~\% in the Fe-N bond length, observed upon the SC, is quite 
a drastic effect.     Therefore, the most remarkable result 
in the present work is the difference observed in the range 
500-720~cm$^{-1}$ between HTHS and PIHS, shown in Fig.~6.    
This demonstrates that {\it the two high-spin states involve 
microscopically different characters.}   
As already mentioned, the expansion of FeN$_6$ cluster 
upon the LTLS~$\rightarrow$ PIHS crossover at 10~K is 
expected to locally create a pressure on the ligand picolylamine.   
However, such pressure is absent in HTHS.    
This additional, local pressure on the ligand picolylamine 
should be responsible for the shifts of lines $b'$, $c$ 
and $b''$ between PIHS and HTHS.     
Note that they are shifted toward higher frequencies in PIHS 
by about 10-20~cm$^{-1}$ than in HTHS .    
These shifts correspond to 3-6~\% increases in the 
effective ``spring constants'' of these vibration modes, 
which is a consequence of the local pressure within the 
picolylamine in PIHS.  
In addition, the appearance of peak $b$ only in PIHS 
is quite remarkable.    This can be attributed to a 
loss of inversion center within the unit cell structure 
in PIHS, since this line has been also observed by Raman 
spectroscopy in all the three states.\cite{taya-PRB}    
(Raman and IR selection rules are exclusive in HTHS and 
LTLS where the unit cell structure has an inversion center.)    
Namely, the unit cell structure in PIHS should be slightly 
deformed from that in HTHS, although the local symmetry 
around Fe$^{2+}$ is kept $O_h$, causing the originally 
IR-inactive $b$ line to appear in PIHS.     
The slight displacement of nnn C atoms, detected by X-ray 
absorption, is a most probable candidate for this deformation.   
The driving force for creating this deformation should be 
the pressure on picolylamine caused by the FeN$_6$ cluster.    

The local deformation and pressure within the picolylamine 
ligands, 
suggested by the present IR and the X-ray absorption 
experiments,\cite{XAS} should be very important in understanding 
the development of PIHS over a macroscopic volume in Fe-pic.    
It should require, as explained in Introduction, a strong, collective 
interaction among the Fe$^{2+}$ ions and a strong coupling 
between the Fe$^{2+}$ electronic state and the lattice.    
Since the X-ray 
absorption results have shown that the Jahn-Teller effect 
is absent in Fe-pic, it is very likely that the deformation 
of picolylamine should play an important role.     
Note that the picolylamine ligands within a [Fe(2-pic)$_3$] 
complex are hydrogen-bonded to those within the 
neighboring complexes.\cite{mikami}   
    Hence the local deformations at the 
picolylamine ligands caused by the expansion of FeN$_6$ 
and the hydrogen bondings connecting the picolylamine 
ligands at neighboring [Fe(2-pic)$_3$] complexes are the 
most likely origin for the collective interaction among the 
Fe$^{2+}$ ions.      To understand such possibility on a 
more microscopic basis, however, much more has to be done 
both experimentally and theoretically.    
In particular, further experiments in the far-IR region 
(at frequencies below 450~cm$^{-1}$) should provide much 
more insight into the photo-induced variations of 
microscopic vibration states, since the vibration modes 
of the FeN$_6$ clusters should be directly observed.   
In addition, lower-frequency modes due to FeN$_6$ and 
picolylamine are expected to be more sensitive 
to collective lattice effects.    
Such experiments are under way.

\section{Conclusions}
We have presented detailed IR absorption 
spectra of the spin-crossover complex Fe-pic in its 
three characteristic states, using single crystal 
samples for the first time.   
To overcome the technical difficulty with using the crystal 
samples, a synchrotron radiation from SPring-8 has been 
used as an IR source.    In addition to the strong 
absorption lines previously observed with powder samples, 
we have newly observed many fine lines.    
Many of the IR lines show 
strong variations upon temperature- and photo-induced SC's.    
Most remarkably, some of the lines below 720 cm$^{-1}$, which 
are due to the skeletal bending vibrations within the 
picolylamine ligands, show clear shifts between HTHS 
and PIHS.    This result demonstrates that HTHS and PIHS 
involve microscopically different characters.     
In particular, the present results suggest that the 
picolylamine ligands receive a strong physical 
pressure upon the photo-induced SC.      
We have conjectured that such local pressure and the resulting 
deformation may have led to the collective, long-range 
interaction needed for the development of macroscopic 
photo-induced phase, 
aided by the hydrogen bondings bridging the Fe-pic complexes.

\section*{Acknowledgements}

We would like to thank M. Matsunami and Y. Kondo for 
technical assistance.   The experiments were performed 
at SPring-8 under the proposal 2002B0220-CS1-np.   
This work is partly supported by Grants-In-Aid from 
the MEXT and by Asahi Glass Foundation.


\begin{table}[b]
\caption{
Assignment of absorption lines to the specific vibration modes 
of Fe-pic.   The letters in the first column denote the 
labels for the observed absorption lines in Fig.~5. 
}
\begin{tabular}{c|c|p{2.in}} 
Peaks$^1$ & Fragments$^2$ & Vibration modes \\ \hline\hline
$a$     & am   &  C-H deformation      \\ \hline
$b, b', b''$   &  am  & skeletal -C-C-N- bending   \\  \hline 
$c$     & pyr  & ring in-plane bending   \\ \hline
$d, d'$ & pyr  & C-H deformation   \\ \hline
$e$     & am   &  C-H (-CH$_2$-) deformation   \\ \hline
$f$     & EtOH & C-H (-CH$_3$) deformation     \\ \hline
$g$     & pyr  & ring in-plane stretching \\ \hline
$h$     & EtOH & -C-OH stretching   \\ \hline
$i$     & pyr  & CH in-plane deformation \\ 
        & am   & -C-N- stretching  \\ \hline
$j$     & am, EtOH & C-H deformation      \\ \hline
$k$     & pyr  & ring in-plane stretch  \\ \hline
$l$     &    &  overtone or coupled.  \\ \hline         

\end{tabular} 
\noindent $^1$ The peak labels refer to those used in Fig.~5. \\
\noindent $^2$ pyr = pyridine ring, am = aminomethyl base, EtOH = ethanol.\\

\end{table}

\begin{figure}
\begin{center}
\caption{
(a) Fe$^{2+}$ and the 2-picolylamine ligands in Fe-pic.  
(b) A 2-picolylamine (2-aminomethyl pyridine) molecule, 
which consists of a pyridine ring and an aminomethyl base.    
(c) Schematic electronic configurations of Fe$^{2+}$ in the 
high-spin ($S$=2) and in the low-spin ($S$=0) states.   
The former has a smaller crystal field splitting and a 
larger Fe-N distance.  
(d) The fraction of Fe$^{2+}$ ions in Fe-pic in the high-spin 
state, $\gamma_{HS}$, as a function of temperature.\cite{taya-thesis}   
$\gamma_{HS}$ was determined by the measured magnetic 
susceptibility.  
}
\end{center}
\end{figure}

\begin{figure}
\begin{center}
\caption{
Optical absorption spectra of Fe-pic at several temperatures.   
(The definition of optical density is explained in the text.)  
Each spectrum is vertically offset by 2.  The shaded areas 
indicate absorption lines that show large 
temperature variations.  Spectral resolution is 4~cm$^{-1}$.
}
\end{center}
\end{figure}

\begin{figure}
\begin{center}
\caption{
Optical absorption spectra of Fe-pic in the 
2100-2500~cm$^{-1}$ range.   The broken curve is 
the spectrum at 140~K, shown in comparison to 
that at 40~K.   The spectral resolution is 4~cm$^{-1}$.
}
\end{center}
\end{figure}

\begin{figure}
\begin{center}
\caption{
The intensity of several absorption lines in Figs.~2 and 3 
as a function of temperature.      Each curve shows the 
intensity integrated over the indicated frequency range, 
normalized by the difference between 40 and 200~K.  
}
\end{center}
\end{figure}

\begin{figure}
\begin{center}
\caption{
Optical absorption spectra of Fe-pic in the low-temperature 
low-spin state (LTLS) 
and in the photo-induced high-spin state (PIHS) at 10~K.    
The spectral resolution is 2~cm$^{-1}$.  
The labels $a$ to $i$ indicate the observed absorption lines, 
and used in the text.  
}
\end{center}
\end{figure}

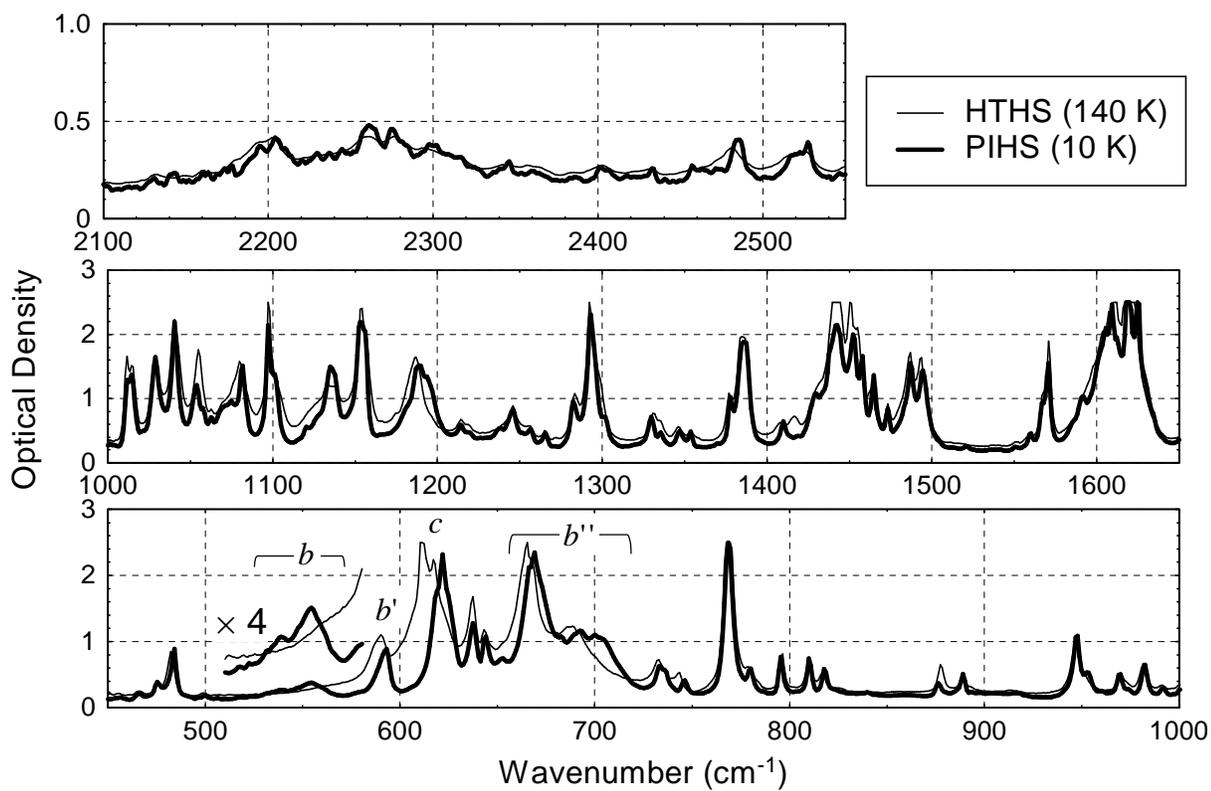
\begin{figure}[h]
\begin{center}
\caption{
Optical absorption spectra of Fe-pic in the photo-induced 
high-spin state (PIHS) at 10~K and in the high-temperature 
high-spin state (HTHS) at 140~K.    
Spectral resolution is 2~cm$^{-1}$.  
The same labels as those in Fig.~5 are indicated for 
several absorption lines which exhibit marked 
differences between the two states.  
}
\end{center}
\end{figure}

\end{document}